\documentclass[12pt,oneside]{article}
\usepackage[english]{babel}
\usepackage{graphics,graphicx,epsfig}
\makeindex
\newcommand{\be}{\begin{equation}}
\newcommand{\ee}{\end{equation}}
\newcommand{\bea}{\begin{eqnarray}}
\newcommand{\eea}{\end{eqnarray}}

\def\asec{$''$ cy$^{-1}$}
\def\ct#1{\cite{#1}}

\def\bb{\bibitem}

\def\eqi{\begin{equation}}
\def\eqf{\end{equation}}
\begin{document}
\begin{titlepage}
\begin{flushright}
\today\\
\end{flushright}
\vspace{.5cm}
\begin{center}
{\LARGE Secular increase of the Astronomical Unit and perihelion
precessions as tests of the Dvali-Gabadadze-Porrati
multi-dimensional braneworld scenario\\} \vspace{0.5cm}
\quad\\
{Lorenzo Iorio\\
\vspace{0.5cm}
\quad\\
Viale Unit\`a di Italia 68, 70125,
Bari, Italy\\ \vspace{0.5cm}
}
\vspace*{0.5cm}

{\bf Abstract\\}
\end{center}

{\noindent \small An unexpected secular increase of the
Astronomical Unit, the length scale of the Solar System, has
recently been reported by three different research groups
(Krasinsky and Brumberg, Pitjeva, Standish). The latest JPL
measurements amount to 7$\pm 2$  m cy$^{-1}$. At present, there
are no explanations able to accommodate such an observed
phenomenon, neither in the realm of classical physics nor in the
usual four-dimensional framework of the Einsteinian General
Relativity. The Dvali-Gabadadze-Porrati braneworld scenario, which
is a multi-dimensional model of gravity aimed to the explanation
of the observed cosmic acceleration without dark energy, predicts,
among other things, a perihelion secular shift, due to Lue and
Starkman,  of $5\times 10^{-4}$ arcsec cy$^{-1}$ for all the
planets of the Solar System. It yields a variation of about $6$ m
cy$^{-1}$ for the Earth-Sun distance which is compatible with the
observed rate of the Astronomical Unit. The recently measured
corrections to the secular motions of the perihelia of the inner
planets of the Solar System are in agreement with the predicted
value of the Lue-Starkman effect for Mercury, Mars and, at a
slightly worse level, the Earth.}
\end{titlepage}
\newpage
\setcounter{page}{1}
\vspace{0.2cm}
\baselineskip 14pt

\setcounter{footnote}{0}
\setlength{\baselineskip}{1.5\baselineskip}
\renewcommand{\theequation}{\mbox{$\arabic{equation}$}}
\noindent

\section{Introduction}
In some of the string theory-inspired multi-dimensional scenarios
the four-dimensional spacetime of our experience is a brane within
a higher-dimensional bulk \ct{BvdBD04}. While the electroweak and
strong interactions are confined in the four-dimensional brane,
gravity can go through the entire bulk experiencing also the extra
dimensions. In most of such theories gravity deviates from its
Newtonian limit at small scales. Instead, a braneworld theory
which yields, among other things, long-range modifications of the
Newton-Einstein gravity has recently been put forth by Dvali,
Gabadadze and Porrati (DGP) \ct{DGP00}. In the DGP model, which
encompasses an extra flat, infinite spatial dimension giving rise
to a five-dimensional Minkowskian bulk, there is a free crossover
parameter $r_0\propto c/H_0$, where $c$ is the speed of light in
vacuum and $H_0$ is the Hubble parameter. Such a scale parameter
is fixed by observations of Type IA Supernov\ae\ to $ r_0\sim 5$
Gpc \ct{DGZ03, LS03}. Beyond $r_0$  gravity becomes
five-dimensional and suffers strong modifications yielding to
cosmological consequences which would yield an alternative to the
dark energy in order to explain the observed acceleration of the
Universe.

Interestingly, in the neighborhood of a central object of mass
$M$, i.e. for\footnote{$G$ is the Newtonian constant of
gravitation.} $R_{\rm g}<< r << r_0,\ R_{\rm g}=2GM/c^2$ where
gravity is four-dimensional and the fifth dimension is naively
hidden, there are also small modifications to the usual
Newton-Einstein gravity which could be soon detectable. Indeed,
Lue and Starkman (LS) derived in \ct{LS03} an extra pericentre
advance \eqi \frac{d\omega}{dt}=\mp \frac{3}{8}\frac{c}{r_0}\eqf
for the orbital motion of a test particle assumed to be almost
circular. It amounts to $\mp5\times 10^{-4}$ arcseconds per
century (\asec\ in the following). The LS precession is an
universal feature of the orbital dynamics of a test particle
because, in this approximation, it is independent of its orbital
parameters. In regard to its sign, it depends on the global
properties of the cosmological phase, i.e. the behavior of the
Universe at cosmological distances. The plus sign is related to
the self-accelerating phase, while the minus sign is due to the
standard FLRW phase \ct{DGZ03}. Thus, it would be possible to get
information on the space-time features at cosmological scales from
local, non-cosmological gravity measurements. The consistency and
the stability of such aspects of DGP gravity have been studied in
\ct{NicRat}.

In view of the recent improvements in the Solar System planetary
ephemerides determination \ct{pIT05, Pit05}, the possibility of
testing the DGP model with local measurements on the inner planets
has been investigated in \ct{Ior05}. It turns out that we are now
at the edge of the observational sensitivity. The secular motion
of the perihelion of Mars is the best known up to now, but a major
source of systematic bias is represented by the orbital
perturbations induced by the asteroids, if not carefully accounted
for. In the case of Mercury, the observational error in
determining the extra-advance of its perihelion is still one order
of magnitude larger than the LS effect. However, as we will show
in Section \ref{pit}, the latest, accurate measurements are not in
disagreement with the DGP predictions.

In Section \ref{auu} we wish to investigate the opportunities to
test DGP gravity offered by the recent discovery of an unexpected
feature of the Solar System dynamics.

\section{The observed secular increase of the Astronomical Unit and its possible explanation in terms
of the DGP gravity}\label{auu}
 An Astronomical Unit (AU), equal to 149597870.691 km, is approximately the mean distance between the
 Earth and the Sun. It is a derived constant and used to indicate distances within the Solar System. Its
 formal definition is the radius of an unperturbed circular orbit a massless body would revolve about the Sun
 in $2\pi/k$ days (i.e., 365.2568983.... days), where $k$ is defined as the Gaussian constant exactly equal
 to 0.01720209895. Since an AU is based on radius of a circular orbit, one AU is actually slightly less than the average distance between the Earth and the Sun which is $\left\langle r \right\rangle=a(1+e^2/2)$ to order $\mathcal{O}(e^2)$, where $a$ and $e$ are the semimajor axis and the eccentricity, respectively.

 Recently, an unexpected secular increase of AU has been measured by Krasinsky and Brumberg \ct{KraBru04} by analyzing the huge wealth of radiometric data of the major planets. The existence of such centennial rate of AU has  been later confirmed also by Standish \ct{Sta05}, who quotes
 \eqi \frac{d {\rm AU}}{dt}=7\pm2\ {\rm m\ cy}^{-1},\eqf and Pitjeva \ct{Sta05} who yields a secular rate of about
 5 m cy$^{-1}$. Note that such estimates and the related uncertainties should be considered rather robust because
 they are not the mere, formal statistical errors but come from many different solutions in which various data sets
 were differently weighted and the parameters of the fit changed.

 In regard to the explanation of such an effect, the discussion in  \ct{KraBru04} rules out many causes both of
 classical and general relativistic origin. In particular, it turned out that the permanent loss of solar mass
 due to electromagnetic radiation and solar wind would induce a centennial rate of AU far smaller than the
 observed one. Also the cosmological expansion of the Universe with uniform mass distribution,
 calculated in the framework of the usual four-dimensional General Relativity cannot explain the measured
 increase of AU. The hypothesis that a secular decrease of the Newtonian gravitational constant
 $\dot G/G=-2\times 10^{-12}$ yr$^{-1}$ can accommodate the measured $d$AU/$dt$ is ruled out by the
 latest measurement  $\dot G/G\sim 10^{-14}$ yr$^{-1}$ by Pitjeva \ct{Pit05}.
 The general relativistic gravitomagnetic Lense-Thirring force was
 not included in the suite of the dynamical force models used in
 the data reduction, but it turns out that it induces for the Earth a secular
 decrease of about -1 m cy$^{-1}$ which cannot be measured with
 the present-day accuracy.
 However, caution is advised because the possibility of some instrumental systematic bias cannot be completely
 ruled out.

 The DGP braneworld model can be used to explain the observed secular increase of AU. Indeed, the shift in the
 Sun-planet distance\footnote{As pointed out in \ct{LS03}, the DGP corrections do not affect the propagation
 of the electromagnetic waves with respect to the general relativistic picture. Thus, no cancellation effects
 analogous to those described in \ct{KraBru04} for General Relativity occur for DGP gravity. } induced by a small
 secular advance of the perihelion  can be approximately evaluated as follows. The position of a planet moving along
 a Keplerian ellipse oscillates between a minimum distance of $r_{\rm min}=a(1-e)$ and a maximum distance of
 $r_{\rm max}=a(1+e)$ over one orbital revolution. If during this temporal interval the ellipse precesses in its plane by
 a small amount $\Delta\omega$, the related shift in the planet's
 position can be approximately written as  $\Delta r\sim ae\Delta\omega$ \ct{Nor00}. The LS
precession, whose magnitude is\footnote{Other generalizations of
the DGP model cannot be tested in Solar System because their
effects are too small \ct{DGZ03}.} $5\times 10^{-4}$ \asec, yields
a centennial rate of 5.7 m cy$^{-1}$ for the Earth ($a=1.00000011$
AU, $e=0.01671022$). In view of a comparison of such a prediction
with the measured values, it must be noted that the uncertainty in
$r_0$ can be evaluated as 0.6 Gpc from the fitting of Supernov\ae\
type IA data \ct{LS03} and the error in the Hubble parameter
\ct{Eli04}. This yields a theoretical range for the DGP shift from
5 to 6.4 m cy$^{-1}$. In view of the present-day accuracy of 2 m
cy$^{-1}$ an uncertainty of 1.4 m cy$^{-1}$ in the theoretical
predictions does not affect the
 comparison with the measured data. The predicted values
 are compatible with the latest measurements. Moreover, the observed plus sign
 points towards the self-accelerating cosmological phase.
 \section{The observed extra-perihelion secular
 advances}\label{pit}
 Pitjeva recently determined a set of corrections to the secular motions of the perihelia of the inner planets of the
 Solar System (Table 3 of \ct{Pit05}) which
 can be used to yield a further, preliminary test of the DGP gravity through the LS
 precessions. In the framework of the EPM2004 ephemerides, more than 317000 observations (1913-2003)
 were used to construct various solutions differing each other for the subset of data used and
the parameters fitted.
 Let us briefly recall that the perihelion of a given planet secularly precesses
 due to the Newtonian N-body interactions
 with the other major bodies of the Solar System, including the largest 301 asteroids and the asteroid ring that
 lies in the ecliptic plane and consists of the remaining smaller asteroids,
 the solar quadrupole mass moment $J_2^{\odot}$
 and  the post-Newtonian
general relativistic gravitoelectric Schwarzschild and
gravitomagnetic Lense-Thirring fields of the Sun. The particular
solution by Pitjeva which we are interested in was constructed by
keeping fixed the Schwarzschild field and $J_2^{\odot}$ to their
reference values and by neglecting the Lense-Thirring
force\footnote{All the other dynamical features of the planetary
motion, including the perturbations from the 301 major asteroids
and the asteroid ring, were included in the models adopted in
\ct{Pit05}. }; corrections to the secular motions of the planetary
perihelia were included in the set of the fitted parameters. Iorio
\ct{Iorsollt} pointed out that such results can yield the first
observational evidence of the solar gravitomagnetic field.

Once the Lense-Thirring precessions are subtracted from the
measured extra-perihelion precessions, it can be checked if the
remaining effects are compatible with the predictions of the DGP
gravity. It turns out that for Mercury and Mars the LS precessions
are compatible with the observed residual precessions:
$(-6.6\times 10^{-3}<5\times 10^{-4}<3.4\times 10^{-3})$ \asec\
for Mercury, $(-4\times 10^{-4}<5\times 10^{-4}<6\times 10^{-4})$
\asec\ for Mars. In the case of the Earth the predicted value of
the LS precession falls slightly outside the measured range whose
upper bound is $3\times 10^{-4}$ \asec. The results for Venus are
rather imprecise due to the very small eccentricity of its orbit
($e=0.00677323$) which makes its perihelion not well defined.

\section{Conclusions}
In this paper we have shown that the DGP multidimensional model of
gravity can be used to explain the recently observed secular
increase of the Astronomical Unit. Indeed, the latest reported
measurement amounts to $7\pm 2$ m cy$^{-1}$, while the magnitude
of the predicted shift is  about 6 m cy$^{-1}$.
In regard to its sign, it is positive and corresponds to the
self-accelerating cosmological phase. At present, there are no
other satisfactorily explanations of the observed secular rate of
the Astronomical Unit, although the impact of possible systematic
errors in the observations should be carefully considered. It
turns also out that the predicted values of the LS perihelion
precessions, based on the DGP model, are compatible with the
recently measured extra-precessions for Mercury, Mars and, at a
slightly worse level, the Earth, although the errors are still
large. Future observations both from ground and from spacecrafts
should allow to improve the accuracy of such tests.

\section*{Acknowledgements}
I thank M. Porrati for interesting discussions and clarifications.

\end{document}